\documentclass[twocolumn,amsmath,amssymb,superscriptaddress,prl,a4paper]{revtex4-1}
\usepackage{graphicx}
\usepackage[paperwidth=212mm,paperheight=297mm,centering,hmargin=1.65cm,vmargin=2.6cm]{geometry}

\usepackage{graphicx}                       
\usepackage{color}                          
\usepackage{hyperref}                       
\usepackage{dsfont}
\usepackage{mathrsfs}
\usepackage{bigints}
\usepackage{bbold}
\usepackage{amsthm}
\usepackage{setspace}

\newcommand{\ket}[1]{\vert#1\rangle}
\newcommand{\bra}[1]{\langle#1\vert}

\newcommand{\beq}{\begin{eqnarray}}
\newcommand{\eeq}{\end{eqnarray}}

\newcommand{\mc}[1]{\mathcal{#1}}
\newcommand{\tr}{\operatorname{tr}}

\begin{document}
\title{Limits to coherent scattering and photon coalescence from\\solid-state quantum emitters}

\author{Jake Iles-Smith}
\email{jakeilessmith@gmail.com}
\affiliation{Department of Photonics Engineering, DTU Fotonik, {\O}rsteds Plads, 2800 Kongens Lyngby, Denmark}
\affiliation{{Photon Science Institute \& School of Physics and Astronomy, The University of Manchester, Oxford Road, Manchester M13 9PL, United Kingdom}}
\author{Dara P. S. McCutcheon}
\affiliation{Department of Photonics Engineering, DTU Fotonik, {\O}rsteds Plads, 2800 Kongens Lyngby, Denmark}
\affiliation{Centre for Quantum Photonics, H.H. Wills Physics Laboratory, University of Bristol, Tyndall Avenue, Bristol BS8 1TL, United Kingdom}
\author{Jesper M\o{}rk}
\affiliation{Department of Photonics Engineering, DTU Fotonik, {\O}rsteds Plads, 2800 Kongens Lyngby, Denmark}
\author{Ahsan Nazir}%
\email{ahsan.nazir@manchester.ac.uk}%
\affiliation{{Photon Science Institute \& School of Physics and Astronomy, The University of Manchester, Oxford Road, Manchester M13 9PL, United Kingdom}}
\date{\today}
\pacs{Valid PACS appear here}

\begin{abstract}
The desire to produce high-quality 
single photons for applications in quantum information science has lead to renewed interest in 
exploring solid-state emitters in the weak excitation regime. 
Under these conditions it is expected that photons are coherently scattered, 
and so benefit from a substantial 
suppression of detrimental interactions between the source and its surrounding 
environment. 
Nevertheless, we demonstrate here that this reasoning is incomplete, 
as phonon interactions continue to play a crucial 
role in determining solid-state emission characteristics even for very weak excitation. 
We find that the sideband resulting from non-Markovian relaxation of the phonon environment is excitation strength independent. It thus 
leads to an intrinsic limit to the fraction of coherently scattered light and to the visibility of two-photon coalescence 
at weak driving, both of which are absent for atomic systems or within simpler Markovian treatments. 
\end{abstract}
\maketitle

In the past few 
years, artificial atoms such as semiconductor quantum dots (QDs) have emerged 
as a leading platform to develop novel photonic sources for applications in 
quantum information science. This interest has been driven in part by a host of experiments establishing that 
QDs exhibit the optical properties of few-level-systems, much like their natural atomic counterparts. 
This includes single photon emission and 
photon antibunching~\cite{Bayer_dots,Santori2002,PhysRevLett.69.3216,PhysRevLett.98.117402,Fedorych12,flagg10}, 
entangled photon emission~\cite{Muller2014,stevenson06}, 
coherent Rabi oscillations~\cite{flagg09,monniello13,ates09}, and resonance 
fluorescence~\cite{ulrich11_short,Konthasinghe2012,Matthiesen2013,C.Dada2016,PhysRevLett.108.093602,Malein2016,Schulte2015,PhysRevLett.114.067401,Kalliakos2016}, 
which has culminated in recent demonstrations of efficiently generated, highly indistinguishable photons~\cite{gazzano13,Ding2016,Somaschi2016,Thoma2016}. 
Moreover, the solid-state nature of QDs offers advantages not shared by atomic systems, 
such as the ease with which they can be optically addressed, larger oscillator strengths, 
and potential embedding into complex photonic structures~\cite{lodahl2015interfacing,Niels_review}. 
However, less advantageous distinctions are also present, 
principally the unavoidable interactions between QD excitonic degrees of freedom and the 
environment provided by the host material~\cite{PhysRevLett.104.017402,PhysRevLett.105.177402,PhysRevB.87.081308,Houel2012,Kuhlmann2013,Kuhlmann2015}. 
These can significantly alter QD optical emission properties~\cite{McCutcheon2013,Iles-Smith:16,hughespopinv,PhysRevLett.114.137401}, 
which typically reduces performance in quantum photonic devices~\cite{PhysRevB.90.035312}.

Recent efforts to suppress the detrimental effects of 
environmental coupling 
in solid-state emitters have renewed interest 
in studying the weak resonant excitation (Heitler) 
limit~\cite{Konthasinghe2012,Matthiesen2013,C.Dada2016,PhysRevLett.108.093602,Malein2016,Schulte2015,PhysRevLett.114.067401}. 
In atomic systems, this regime is dominated by elastic (coherent) scattering of photons, 
with the proportion of coherent emission approaching unity as the driving strength is reduced~\cite{carmichael1998statistical}. 
As the population excited within the emitter then becomes very small, it is expected that in solid-state 
systems 
the effects of any environmental interactions will correspondingly be suppressed, 
such that the emitted photon coherence times may become extremely long.  


Here, 
we demonstrate that this intuition is incomplete as  
phonon interactions remain a vital consideration for QDs in the weak excitation regime, 
despite the vanishing dot population. 
Specifically, we show that the sideband resulting from non-Markovian relaxation of the phonon environment is excitation strength independent, unlike the well-studied Markovian phonon contribution~\cite{PhysRevLett.104.017402,PhysRevLett.105.177402}. It thus 
leads
to an intrinsic sub-unity limit 
to the fraction of coherently scattered light from a solid-state emitter,  
even 
in the absence of charge fluctuations 
and no matter how weak the driving.
%
This is in clear contrast to the atomic case, 
constituting a novel regime of semiconductor quantum optics. 
It is also of direct practical importance,  
for example to 
light--matter coupling schemes that rely on the coherent scattering of photons 
with high efficiency~\cite{Hu2008,hu2014saturation}.
Furthermore, we show that 
the impact of the phonon relaxation process 
can be even more pronounced in two-photon interference experiments, resulting in a substantial suppression of 
the 
photon coalescence visibility on picosecond timescales, which is  
exacerbated when accounting for the inevitable detector temporal response. 
This leads to a non-monotonic dependence of the visibility on driving strength that is 
unexpectedly optimised at intermediate rather than very weak excitation. 

\emph{Model---}We model the QD as a two-level-system~\cite{bimberg1999quantum,zrenner02,1367-2630-12-11-113042,PhysRevLett.104.017402,PhysRevLett.105.177402}, having upper (single exciton) state $\ket{X}$ of frequency $\omega_X$    
and ground state $\ket{0}$,   
under continuous-wave (CW) excitation with Rabi frequency 
$\Omega$ and detuning $\delta$. Generalising to pulsed excitation is possible by introducing a time dependent Rabi frequency in analogy with Ref.~\cite{1367-2630-12-11-113042}. 
The electromagnetic and vibrational environments are treated 
as two separate reservoirs of harmonic oscillators.
Within a frame rotating 
at the laser frequency $\omega_l$,   
the Hamiltonian may be written ($\hbar = 1$):
\begin{align}\label{Hfull}
H&=\delta\sigma^\dagger\sigma+ \frac{\Omega}{2}\sigma_x +\sum\limits_k \nu_kb_k^\dagger b_k + \sum\limits_m\omega_m a^\dagger_ma_m\nonumber\\
&+ \sigma^\dagger\sigma\sum\limits_k g_k(b_k^\dagger+b_k) +\sum\limits_m f_m \sigma^\dagger a_m e^{i\omega_l t}+ \text{h.c.},
\end{align}
%
where $\sigma^\dagger = \ket{X}\bra{0}$, $\sigma_x = \sigma^\dagger + \sigma$, and we have applied a rotating-wave approximation to the driving field. 
Each phonon mode is characterised by a creation (annihilation) 
operator $\smash{b^\dagger_k}$ ($b_k$), frequency $\nu_k$, and couples to the system with strength $g_k$.
Similarly, the 
electromagnetic environment modes 
are defined by creation (annihilation) operators $\smash{a_m^{\dagger}}$ ($a_m$),  
frequencies $\omega_m$, and couplings $f_m$. 

The interactions between the QD and the two harmonic environments are  
determined by their spectral densities. For the phonon environment, 
we take the standard form~\cite{nazir2008photon,PhysRevLett.105.177402,PhysRevLett.104.017402} 
$J_{\rm{PH}}(\nu)= \sum_kg_k^2\delta(\nu-\nu_k)=\alpha\nu^3\exp(-\nu^2/\nu_c^2)$, where $\alpha$ is the 
system--environment coupling strength 
and $\nu_c$ is the phonon cut-off frequency, 
the inverse of which approximately specifies the phonon bath relaxation timescale 
and is directly related to the ratio of 
the QD size $d$ and the speed of sound $v$~\cite{nazir2008photon}. 
The situation is simplified for the electromagnetic environment; for a QD in a bulk 
medium~\footnote{Or, for example, a QD placed in the centre of a photonic crystal waveguide 
within the Purcell regime.}, the local density of states of the electromagnetic field does not vary 
appreciably over the relevant QD energy scales. 
This allows us to assume the spectral density to be flat~\cite{carmichael1998statistical,McCutcheon2013}, 
$J_{\rm{EM}}(\omega)=\sum_m\vert f_m\vert^2\delta(\omega-\omega_m)\approx2\gamma/\pi$, where $\gamma$ is the spontaneous emission rate. 

The dynamics generated by Eq.~(\ref{Hfull}) are not in general amenable to exact solutions.
However, in regimes relevant to QD systems we may derive a very accurate master equation (ME) for the reduced 
state of the QD using the polaron formalism~\cite{1367-2630-12-11-113042,PhysRevLett.113.097401,Roy2011X,nazir2015modelling}, 
which is valid beyond the standard limit of weak 
QD-phonon coupling~\cite{1367-2630-12-11-113042,nazir2008photon}. As we shall see, our treatment, 
though Markovian in the polaron representation, still retains non-Markovian processes for operators evaluated in the original representation. 
This makes it particularly well suited to probing novel phonon effects in QD optical emission properties, as it allows us to 
draw a formal connection between the QD dynamics (generated by the ME) and the characteristics of the emitted electromagnetic field 
via the quantum regression theorem in the usual way, though without imposing 
restrictions to Markovian or weak-coupling regimes 
between the QD and phonons~\cite{mccutcheon2015optical}. 
This is especially important in the weak driving limit, where we shall show that  
non-Markovian relaxation of the phonon environment has a particularly pronounced effect.  

To derive the ME we apply a polaron transformation 
to Eq.~(\ref{Hfull}), defined through $H_{\rm P}=\mathcal{U}_{\rm P}H\mathcal{U}_{\rm P}^{\dagger}$ where 
$\mathcal{U}_{\rm P} = \ket{0}\bra{0} + \ket{X}\bra{X} B_{+}$, 
with $\smash{B_{\pm} = \exp[\pm\sum_k g_k ( b_k^\dagger-b_k)/\nu_k]}$. 
This removes the linear QD-phonon coupling term, 
resulting in a transformed Hamiltonian that may be written as $H_{\rm P}=H_{\rm P}^{0}+H_{\rm P}^{\rm I}$, with 
\begin{align}\label{eq:pol_ham}
H_{\rm P}^{0}&=\tilde\delta\sigma^\dagger\sigma+ \frac{\Omega_r}{2}\sigma_x +\sum_k \nu_kb_k^\dagger b_k + \sum_m\omega_m a^\dagger_ma_m,\\
H_{\rm P}^{\rm I}&=\frac{\Omega}{2}\sigma^{\dagger}(B_+-B) +\sum_m f_m \sigma^\dagger B_+a_m e^{i\omega_l t}+ \text{h.c.}
\label{HPI}
\end{align}
Here 
$\tilde\delta = \delta - \sum_k g_k^2/\nu_k$ is the phonon shifted detuning  
and $\Omega_r = \Omega B$ is the 
renormalised Rabi frequency, with $B=\tr(B_{\pm}\rho_B)$ the average displacement of the phonon environment. For a thermal state of the phonons we have 
$\smash{\rho_B=\exp[-\beta\sum_k \nu_kb_k^\dagger b_k]/\tr_B(\exp[-\beta\sum_k \nu_kb_k^\dagger b_k])}$ with temperature $T=(k_B\beta)^{-1}$, and we find 
$B=\exp[-\frac{1}{2}\int_0^\infty\nu^{-2}J_{\rm{PH}}(\nu)\coth(\beta\nu/2)d\nu]$. 
Tracing out the environments within the Born-Markov approximations~\cite{breuer2007theory}, we obtain a polaron frame ME 
that is second-order in $H_{\rm P}^{\rm I}$ but non-perturbative in the original QD-phonon coupling. 
For $\tilde{\delta}=0$ the ME can be written~\cite{supplement} 
 \begin{equation}
 \dot{\rho}_{\rm P}(t)=-\frac{i\Omega_r}{2}\left[\sigma_x,\rho_{\rm P}(t)\right] + \mathcal{K}_{\rm{PH}}[\rho_{\rm P}(t)] + \mathcal{K}_{\rm{EM}}[\rho_{\rm P}(t)], 
 \label{rhodot}
 \end{equation}
%
where $\mathcal{K}_{\rm{EM}}[\rho_{\rm P}(t)]=\frac{\gamma}{2}\left(2\sigma\rho_{\rm P}(t)\sigma^\dagger -\left\{\sigma^\dagger\sigma,\rho_{\rm P}(t)\right\}\right)$ 
arises from the second term in Eq.~({\ref{HPI}}) and 
describes spontaneous emission, 
with $\rho_{\rm P}$ the polaron frame reduced QD density operator. 
Markovian dissipative processes due to phonons are encoded in the superoperator $\mathcal{K}_{\rm{PH}}[\rho_{\rm P}(t)]$ which originates from the 
first term in Eq.~({\ref{HPI}})~\cite{supplement}. Though the form of $\mathcal{K}_{\rm{PH}}[\rho_{\rm P}(t)]$ can in general be rather complicated, 
it is evident that the influence of these {\it Markovian} phonon terms becomes negligible as $\Omega/\gamma \to 0$ 
and the weak-driving (Heitler) regime is approached, in line with conventional expectations. 

\emph{Photon emission---}
We shall now see how the polaron formalism also allows for non-Markovian phonon processes 
to be readily included into the emitted field characteristics. Crucially, we shall find that this influence {\it does not} disappear as $\Omega/\gamma\rightarrow0$, unlike the standard Markovian one. 
We consider the electric field operator in the Heisenberg picture, which neglecting polarisation may be written 
$
\hat{\bm{E}}(t)= \hat{\bm{E}}^{(+)}(t)+ \hat{\bm{E}}^{(-)}(t),
$ %
with positive frequency component 
$\hat{\bm{E}}^{(+)}(t) =[\hat{\bm{E}}^{(-)}(t)]^\dagger= \sum_m \mathcal{E}_m \hat{a}_m(t)$
where $\mathcal{E}_m$ is the electric field strength. 
Using the formal solution of the Heisenberg equation for $a_m(t)$ we may write 
%
%
$
\hat{E}^{(+)}(t)=
-i\sum_m\int_0^t dt^\prime\mc{E}_m f_m\tilde{\sigma}(t^\prime)B_{-}(t^\prime) e^{i\omega_m (t^\prime -t)}$,
where $\tilde\sigma(t) = \sigma(t) \mathrm{e}^{-i\omega_l t}$, and we have omitted 
the free field contribution $\sum_m\mc{E}_m \hat{a}_m(0) e^{-i\omega_m t}$, 
valid when taking expectation values 
assuming a free field in the vacuum state. 
Using, as before, the fact that the coupling between the emitter and field does not vary 
appreciably over energy scales relevant to the QD, we then obtain 
\begin{equation}\label{eq:TLS_field}
\begin{split}
\hat{E}^{(+)}(t)&\rightarrow -i\mc{E} \sqrt{\frac{\pi\gamma}{2}}\tilde{\sigma}(t)B_{-}(t), 
\end{split}
\end{equation}
%
which we see contains the phonon 
displacement operator $B_-(t)$. 
This results from the transformation to the polaron frame, and captures 
the relaxation of the vibrational environment when a photon is scattered.
This is an inherently non-Markovian process, 
occurring on a 
timescale set by $1/\nu_c\propto d/v\sim 1~\mathrm{ps}$~\cite{mccutcheon2015optical}. 
Interestingly, such non-Markovian 
effects are not evident in the QD populations, studied e.g.~in Ref.~\cite{1367-2630-12-11-113042}, as the relevant operators commute with the polaron transformation. 

The impact of the phonon relaxation process can 
be observed, 
however, in the 
steady-state intensity spectrum of light emitted from the QD, where its 
picosecond timescale 
translates to a broad meV-scale feature. 
The spectrum is related to the field operators through 
the Wiener-Khinchin theorem, 
$S(\omega)=\lim_{t\to\infty}\operatorname{Re}[\int_0^\infty \langle\hat{E}^{(-)}(t)\hat{E}^{(+)}(t+\tau)\rangle e^{i(\omega-\omega_l)\tau}d\tau]$~\cite{carmichael1998statistical}.
Using Eq.~({\ref{eq:TLS_field}}) we find 
$S(\omega)\propto\operatorname{Re}[\int_0^\infty g^{(1)}(\tau)e^{i(\omega-\omega_l)\tau}d\tau]$, 
with $g^{(1)}(\tau)=\lim_{t\rightarrow\infty}\langle{\sigma^\dagger(t)B_{+}(t)\sigma(t+\tau)B_{-}(t+\tau)}\rangle$~\cite{Roy-Choudhury:15,PhysRevB.92.205406,roy2015spontaneous}. 
The level of coherent scattering is determined by the long time limit of the correlation function, 
$\smash{g^{(1)}_{\rm coh}=\lim_{\tau\rightarrow\infty}[g^{(1)}(\tau)]}$, 
and the incoherent emission spectrum 
as $S_{\rm inc}(\omega)=\operatorname{Re}[\int_0^\infty (g^{(1)}(\tau)-g^{(1)}_{\rm coh})e^{i(\omega-\omega_l)\tau}d\tau]$. 
For typical QD systems, $g^{(1)}(\tau)$ contains two quite distinct timescales; the aforementioned picosecond  
timescale 
associated with phonon relaxation, 
and a much longer ($\sim$~ns)  
timescale corresponding to 
photon emission. 
This allows us to factorise the correlation function into short- and long-time contributions, such that 
$\smash{g^{(1)}(\tau) = G(-\tau)g_{0}^{(1)}(\tau)}$, 
where $\smash{g^{(1)}_{0}(\tau)=\lim_{t\rightarrow\infty}\langle\sigma^\dagger(t)\sigma(t+\tau)\rangle}$ can be calculated from 
Eq.~({\ref{rhodot}}) using the (Markovian) 
regression theorem, while 
$G(\tau) = \langle B_{+}(\tau) B_{-}\rangle=B^2\exp[\int_0^\infty \nu^{-2}J_{\rm{PH}}(\nu)(\coth\left(\beta\nu/2\right)\cos\nu\tau - i\sin\nu\tau)d\nu]$ 
describes short-time 
phonon bath relaxation. 

\begin{figure}[t]
\center
\includegraphics[width =0.485\textwidth]{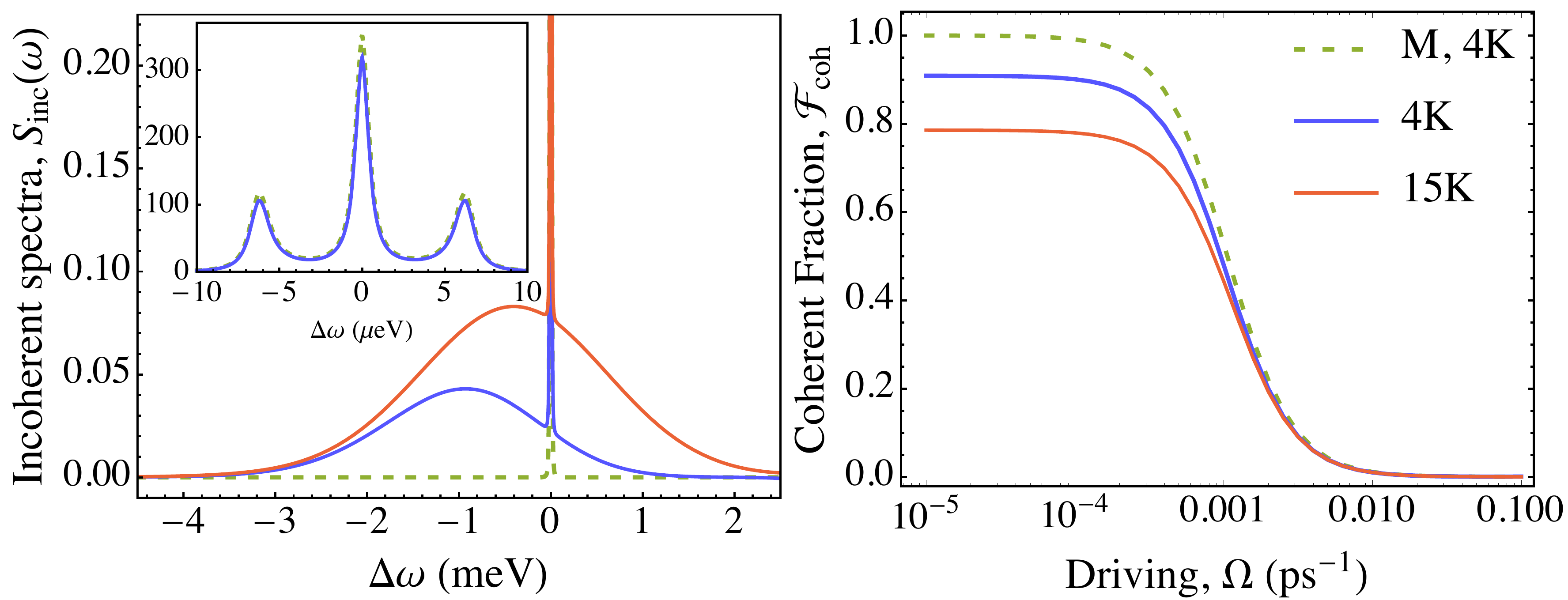}
\caption{Left:~Incoherent emission spectrum for a CW driven QD, demonstrating the presence of the broad phonon 
sideband captured by the non-Markovian theory (solid), when compared to the Markovian expression (dashed). 
The inset shows a zoom around the QD resonance at $4$~K. Right:~Fraction of coherent scattering as a function of driving strength 
within the non-Markovian theory (solid) and the Markovian approximation (M, dashed). Parameters: $\Omega = 0.01$~ps$^{-1}$, $\gamma^{-1} = 700$~ps, 
with the phonon environment characterised by $\alpha = 0.03$~ps$^{2}$, $\omega_c = 2.2$~ps$^{-1}$, and temperatures as indicated.}
\label{SpectraANDF}
\end{figure}

\emph{Spectra---}The impact of these non-Markovian phonon processes is illustrated 
in Fig.~\ref{SpectraANDF}~(left), where 
we plot the incoherent spectrum including (solid lines) and excluding (dashed lines) the short-time phonon contribution $G(\tau)$. 
The inset shows a zoom around $\Delta\omega=\omega-\omega_l=0$, where both approaches capture the Mollow triplet, while 
only the non-Markovian theory captures the broad sideband visible on the scale of the main plot. 
Such sidebands have 
been observed in resonance fluorescence experiments 
on QDs~\cite{Konthasinghe2012,PhysRevLett.108.093602,Matthiesen2013,C.Dada2016}, and previously studied 
theoretically using a non-Markovian regression theorem~\cite{mccutcheon2015optical}. 
Not only does our approach provide a simple method for capturing these 
contributions (i.e.~one that does not rely on non-Markovian extensions to the regression theorem), 
but it also allows us to easily separate the phonon sideband and 
direct emission into independent 
contributions, since 
$S(\omega)=\operatorname{Re}[\int_0^\infty (G(-\tau)-B^2)g^{(1)}_0(\tau)e^{-i(\omega-\omega_l)\tau}d\tau]
+B^2\operatorname{Re}[\int_0^\infty g^{(1)}_0(\tau)e^{-i(\omega-\omega_l)\tau}d\tau]$. 
Here, the first term corresponds to the 
sideband emission, and we make 
use of 
the fact that $G(-\tau)\rightarrow B^2$ after approximately $1$~ps, on which timescale 
$\smash{g^{(1)}_0(\tau)}$ is almost static. Integrating the spectrum over all frequencies we find that the fraction of
power emitted via the phonon sideband is given by $(1-B^2)$. 
Thus, even at $T=0$ the sideband constitutes $(1-B^2)\approx 7\%$ of the total emission, 
rising to $9.1\%$ at $T=4$~K and $22.5\%$ at $T=15$~K, consistent 
with experimental observations~\cite{Matthiesen2013,C.Dada2016}. In the time domain the sideband 
corresponds to a rapid (ps) decrease of the $g^{(1)}$ fringe visibility to $B^2\approx 90.9\%$ at $4$~K, 
which is also in accord with experiments performed in the Heitler limit~\cite{PhysRevLett.108.093602}.

Our formalism also reveals important new physics. It is apparent from the expression for $g^{(1)}(\tau)$ 
that the fraction of light emitted through the phonon sideband is independent of the laser excitation conditions. 
This has particularly significant implications at weak driving, where it affects 
the balance of coherent and incoherent emission. 
For atomic or Markovian systems under very weak excitation, light incident on the emitter 
scatters predominantly elastically, maintaining phase coherence with the driving field. 
However, we now see that for QDs, when accounting for non-Markovian phonon bath relaxation, 
some fraction of the light is \emph{always} emitted incoherently through the sideband 
regardless of the driving strength. Hence the fraction of coherently scattered light is 
reduced below unity at weak driving, as can be seen in Fig.~\ref{SpectraANDF} (right), 
leading to an 
intrinsic limit to the level of coherent scattering 
from such a solid-state photonic system that worsens as temperature is raised. 
Although we have formulated our treatment for the case of CW excitation, the independence of the sideband contribution on driving conditions implies that it 
will also be present for pulsed excitation, 
which is consistent with experiment too~\cite{C.Dada2016}.


Though the limit to coherent scattering could be overcome by filtering out the sideband, 
in this case, the non-Markovian 
phonon relaxation process still imposes an intrinsic operational limitation 
as it leads now to a loss in efficiency of $1-B^2$. A trade-off then exists between the level of coherent scattering 
and the source efficiency. 
This reduction in efficiency may to some extent be mitigated by Purcell enhancement 
for a QD coupled to a narrow cavity mode~\cite{Grange2015}, though 
this too cannot be done arbitrarily, as the cavity can only enhance the $B^2$ fraction of light falling within its linewidth. 

\begin{figure}[t]
\center
\includegraphics[width =0.48\textwidth]{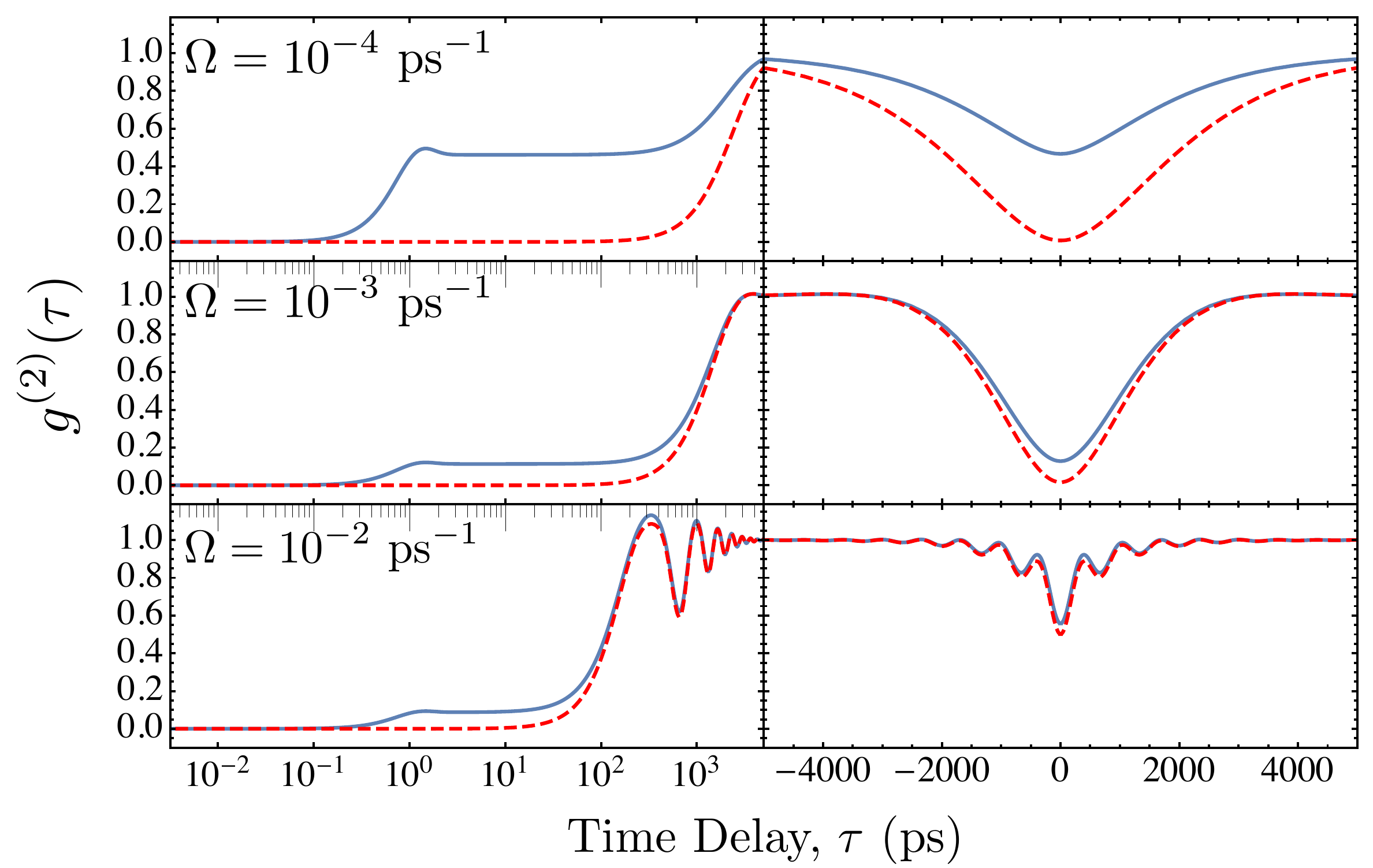}
\caption{HOM second-order correlation function before (left, log-scale) and 
post (right) detector convolution, 
including (solid) and without (dashed) non-Markovian phonon relaxation processes. 
At weak excitation, the convolved HOM-dip is significantly 
shallower when including non-Markovian relaxation. 
Parameters: as Fig.~\ref{SpectraANDF} with $T=4$~K.
} 
\label{fig:fig2}
\end{figure}

\emph{Two photon coalescence---}Given that phonon relaxation acts  
to reduce first-order coherence, 
it is natural to ask whether it also affects the visibility 
of two-photon interference as measured in a Hong-Ou-Mandel (HOM) 
experiment. 
We consider the steady-state intensity 
correlation function 
$g^{(2)}(\tau)=\lim_{t\rightarrow\infty}\langle \hat{E}_3^{-}(t)\hat{E}_4^{-}(t+\tau)\hat{E}_4^{+}(t+\tau)\hat{E}_3^{+}(t)\rangle/
(\langle \hat{E}_3^{-}\hat{E}_3^{+}\rangle\langle \hat{E}_4^{-}\hat{E}_4^{+}\rangle)$ as measured by an unbalanced Mach--Zehnder interferometer, 
with detected output fields $\hat{E}_3(t)$ and $\hat{E}_4(t)$. 
We calculate $g^{(2)}(\tau)$ in the same manner 
as the first order correlation function, where 
phonon operators enter again via Eq.~({\ref{eq:TLS_field}})~\cite{supplement}.

In contrast to two-photon interference experiments using 
pulsed excitation, for CW systems the time resolution of the photon detectors becomes an important 
consideration~\cite{PhysRevLett.114.067401}.  
For example, for an ideal detector with perfect 
time resolution, complete coalescence 
[$g^{(2)}(0)=0$] may be observed at zero time delay regardless of the spectral indistinguishability 
of the incident photons~\cite{PhysRevA.77.042323,PhysRevLett.93.070503,Legero1}; 
the detectors being unable to distinguish frequency or phase differences between the two. 
However, the more distinguishable the photons are, the smaller the time window over which $g^{(2)}(\tau)\approx0$~\cite{PhysRevLett.114.067401}. 
As such, photon detectors with realistic response times will not resolve two-photon interference for sufficiently distinguishable photons. 

The influence of phonon bath 
relaxation on two-photon coalescence is seen in Fig.~\ref{fig:fig2} (left), where we plot 
$g^{(2)}(\tau)$ below saturation ($s=\sqrt{2}\Omega/\gamma\approx0.1$, top), at saturation ($s\approx1$, middle), and above ($s\approx10$, bottom), 
assuming perfect detectors.  
Phonon bath relaxation manifests as 
a sharp short-time feature 
around $\tau=0$, 
clearly apparent in the non-Markovian theory (solid curves), 
and 
particularly pronounced at weak excitation. 
In contrast, 
the Markovian theory (dashed curves) predicts  
much slower dynamics at weak driving strengths, a consequence of 
the vanishing phonon influence within this approach. 
To account for non-ideal detectors, we convolve the correlation 
function with a Gaussian response function 
$R(x) = (2/\delta\tau)\sqrt{\log2/\pi}\exp[-4\log2x^2/\delta\tau^2]$ of  
full width at half maximum $\delta\tau = 400$~ps~\cite{PhysRevLett.100.207405}. 
This gives the intensity correlation function as measured in a realistic experiment, 
and the result is shown in Fig.~\ref{fig:fig2} (right). 
We see that the convolution washes out the detailed features associated to phonon bath relaxation as the detectors are unable to resolve the dip at zero time-delay, reducing the effective visibility of two-photon interference.

\begin{figure}[t]
\center
\includegraphics[width =0.4\textwidth]{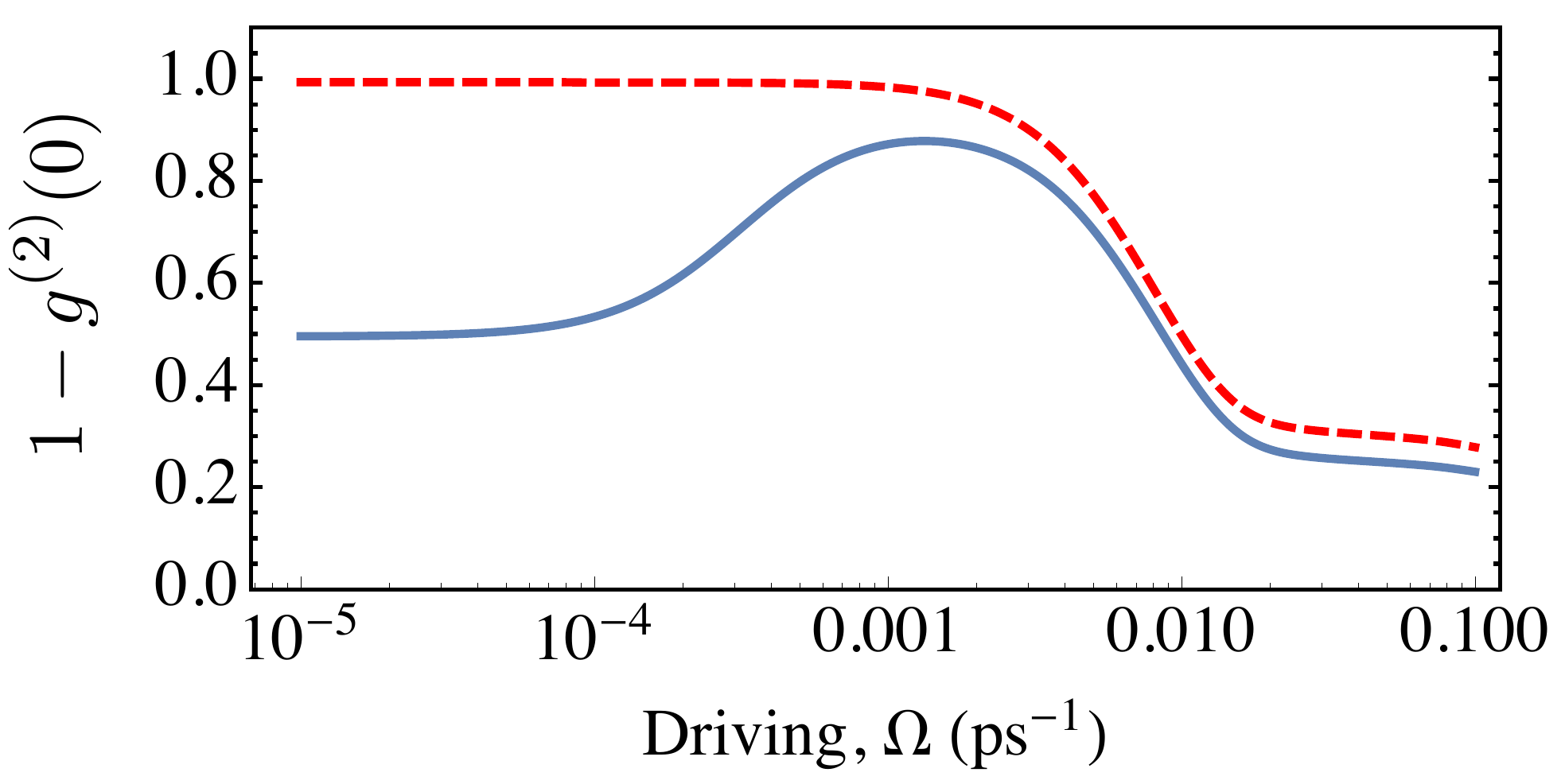}
\caption{HOM-dip depth post detector convolution as a function of driving strength for the Markovian (dashed) and non-Markovian (solid) 
theories. Parameters as in Fig.~\ref{fig:fig2}.}
\label{fig:fig3}
\end{figure}

This is highlighted by Fig.~\ref{fig:fig3}, where the measured dip depth post convolution 
is shown as a function of driving strength. At weak driving (below saturation), the Markovian theory predicts 
perfect interference as the photons are then 
unaffected by phonons, and share the same coherence properties as the driving field. 
Accounting for phonon relaxation, however, the coalescence visibility 
drops dramatically to $1-g^{(2)}(0)\approx0.5$. This reduction occurs 
due to the long timescale associated with optical scattering processes in the weak driving limit, 
which allows the phonon environment to relax fully between scattering events. 
As the driving strength is increased, so too is the rate at which photons are scattered, 
and thus the visibility 
improves. 
Above saturation ($s\approx10$ for $\Omega\sim10^{-2}$~ps$^{-1}$), the correlation function develops Rabi oscillations, 
which cannot be fully resolved by the detector when $\delta\tau\sim\Omega^{-1}$, resulting in a drop in visibility. 

An important implication of these results is that for CW driven solid state emitters with realistic detectors, the weak excitation 
regime is not optimal for generating 
indistinguishable photons. 
Instead, this lies near the onset of strong driving (i.e.~around saturation), 
where in fact the level of incoherent scattering can be larger than the coherent contribution. 
This is in marked contrast to atomic systems, where increasing the fraction of coherent scattering always improves the visibility.
We note that two-photon interference has been observed experimentally with QDs from below to above saturation~\cite{PhysRevLett.114.067401,Kalliakos2016,Malein2016}, thus our predictions should be testable 
using unfiltered emission. 

\emph{Summary---}We have shown that non-Markovian phonon bath 
relaxation processes in driven QDs are excitation strength independent. They thus 
have a profound impact on the level of coherently and incoherently scattered light, limiting 
the coherent fraction to values of $\approx 90\%$ at $T=4~\mathrm{K}$. Moreover, 
when accounting for any realistic detector response time, these short-time phonon relaxation processes 
act to decrease two-photon HOM interference visibilities. 
These results have important implications for numerous quantum technology applications where an efficient 
source of coherently scattered photons is needed as a resource~\cite{Hu2008,hu2014saturation}. 
%
We stress that although our calculations have been formulated in the context of QDs, 
the results are expected to be applicable to a variety of solid-state emitters, including 
nitrogen vacancy centres~\cite{PhysRevLett.85.290}, superconducting qubits, and dye molecules embedded in crystalline lattices~\cite{1367-2630-13-8-085009,Polisseni:16}. 

\begin{acknowledgements}

\emph{Acknowledgements---}JIS and JM acknowledge support from the Danish Research Council (DFF-4181-00416) and Villum Fonden (NATEC Centre). AN is supported by University of Manchester and the Engineering and Physical
Sciences Research Council.
DPSM acknowledges support from a Marie-Curie Individual Fellowship and 
project SIQUTE (contract EXL02) of the European Metrology Research Programme (EMRP). 
The EMRP is jointly funded by the EMRP participating countries within EURAMET and the European Union.

\end{acknowledgements}

\providecommand{\noopsort}[1]{}\providecommand{\singleletter}[1]{#1}%

\clearpage

\widetext
\section{Supplemental Material}

In this supplement we outline the mathematical formalism used in the main text.
We first derive the polaron master equation describing the dynamical evolution of a driven quantum dot (QD), subject to 
interactions with both phonon and electromagnetic environments. 
We then proceed to link this description to the optical properties of the QD, deriving expressions for the first and second-order optical emission correlation functions which also account for non-Markovian phonon environment relaxation processes.

\section{The Polaron Master Equation}
The starting point is the Hamiltonian describing a quantum dot (QD) driven by a classical laser field with frequency $\omega_l$ as given in the main text,
\begin{equation}
H = \delta\sigma^\dagger\sigma + \frac{\Omega}{2}\sigma_x + \sigma^\dagger\sigma\sum\limits_k g_k(b^\dagger_k + b_k) + \sum\limits_m \left(f_m \sigma^\dagger a_m e^{i\omega_l t} + \text{h.c.}\right)
+\sum\limits_k \nu_k b^\dagger_k b_k + \sum\limits_m \omega_m a^\dagger_m a_m.
\end{equation}
Here, the Hamiltonian is written in a frame rotating with respect to the laser frequency $\omega_l$, with the laser and QD transition detuned by $\delta = \omega_X - \omega_l$.
To model the dynamical properties of the QD, we start by applying a unitary polaron transformation to the above Hamiltonian~\cite{1367-2630-12-11-113042s}, allowing us to derive a master equation valid outside the weak exciton-phonon coupling regime. 
In this transformed representation, the Hamiltonian is given by $H_{\rm P} = \mathcal{U}_{\rm P} H \mathcal{U}_{\rm P} ^\dagger = H_S + H_I^{{\rm {\rm PH}}} + H_I^{{\rm EM}}+H_B$, where 
$$\mathcal{U}_{\rm P}= \ket{0}\bra{0} + \ket{X}\bra{X} B_+,$$ 
with $B_{\pm}=\exp( \pm \sum_k g_k (b_k^\dagger - b_k)/\nu_k)$, $H_S = (\delta+R)\sigma^\dagger\sigma + \frac{\Omega_r}{2}\sigma_x$, and $H_B=\sum\limits_k \nu_k b^\dagger_k b_k + \sum\limits_m \omega_m a^\dagger_m a_m$. 
Notice that the Rabi frequency, $\Omega_r = \Omega B$, has been renormalised by the average displacement of the phonon environment $B = \text{tr}_B(B_{\pm}\rho_B^{{\rm {\rm PH}}})$, where $\rho_B^{{\rm {\rm PH}}} =\mathcal{Z}^{-1} \exp(-\beta\sum_k\nu_k b_k^\dagger b_k)$ is the Gibbs state of the phonon environment, with $\mathcal{Z} = \text{tr}_B\left[\exp(-\beta\sum_k\nu_k b_k^\dagger b_k)\right]$, 
and $\beta$ the inverse temperature. 
In addition, the QD resonance has been shifted by $R =\sum_k g_k^2/\nu_k$ as a consequence of the interaction between the QD and the phonon environment. 
For the rest of this work we shall assume that the QD is driven on resonance with the polaron shifted splitting, that is, $\delta +R  = \omega_X^\prime - \omega_l= 0$ with $\omega^\prime_X = \omega_X + R$. 

The interaction terms in the transformed frame take the form
\begin{equation}
H_I^{{\rm {\rm PH}}} = \frac{\Omega}{2}\left(\sigma_x B_x + \sigma_yB_y\right)~~~\text{and}~~~ H_I^{{\rm EM}} = \sum\limits_m f_m\sigma^\dagger B_+ a_m e^{i\omega_l t}+f^\ast_m\sigma B_- a^\dagger_m e^{-i\omega_l t},
\end{equation}
with $B_x = (B_+ + B_- - 2B)/2$ and $B_y = i(B_+ - B_-)/2$. Notice that the interaction between the QD and the electromagnetic field has now obtained displacement operators which act on the phonon environment, while the phonon interaction term contains only system and phonon operators.

To describe the dynamics of the reduced state of the QD, $\rho_S(t)$, we shall treat the interaction Hamiltonian $H_I = H_I^{{\rm {\rm EM}}} + H_I^{{\rm {\rm PH}}}$ to 2$^{nd}$-order using a Born-Markov master equation, which in the interaction picture takes the form~\cite{breuer2007theorys}:
\begin{equation} 
\frac{\partial\tilde\rho_S(t)}{\partial t}=-\int\limits_0^\infty d\tau ~\text{tr}_B\left[H_I(t),\left[H_I(t-\tau), \tilde\rho_S(t)\otimes\rho_B^{{\rm EM}}\otimes\rho_B^{{\rm {\rm PH}}}\right]\right],
\end{equation}
where in the Born approximation we factorise the environmental density operators {\it in the polaron frame} such that they remain 
static throughout the evolution of the system. Note that correlations may be generated between the system and the phonon environment in the original representation. 
We shall assume that in the polaron frame the phonon environment remains in the Gibbs state defined above, while the electromagnetic environment remains in its vacuum state $\rho_B^{{\rm EM}} = \bigotimes_m \ket{0_m}\bra{0_m}$.
Since the trace over the chosen states of the environments removes terms linear in creation and annihilation operators, we may split the master equation into two separate contributions corresponding to the phonon and photon baths respectively~\cite{mccutcheon2015opticals},
\begin{equation}
\frac{\partial\tilde\rho_S(t)}{\partial t}= \mathcal{K}_{{\rm {\rm PH}}}[\tilde\rho_S(t)] + \mathcal{K}_{{\rm EM}}[\tilde\rho_S(t)].
\end{equation}
In the subsequent sections we shall analyse each of these contributions in turn.

\subsection{The phonon contribution}
To derive the contribution from the phonon environment, we follow Ref.~\cite{1367-2630-12-11-113042s}.
We start by transforming into the interaction picture with respect to the Hamiltonian $H_0=\frac{\Omega_r}{2}\sigma_x + \sum_k\nu_k b_k^\dagger b_k + \sum_m \omega_m a_m^\dagger a_m$. 
Using this transformation, the interaction Hamiltonian takes the form:
\begin{equation}
H^{{\rm {\rm PH}}}_I = \frac{\Omega}{2}\left(\sigma_x(t) B_x(t) + \sigma_y(t) B_y(t)\right).
\end{equation}
Here $B_x (t) =\frac{1}{2} (B_+(t) + B_-(t) - 2B)$ and $B_y(t) = \frac{i}{2}(B_+(t)-B_-(t))$, where $B_\pm(t) = \exp[\pm\sum_k\frac{g_k}{\nu_k} (b^\dagger_k e^{i\nu_kt} -b_ke^{-i\nu_kt})]$. The time evolution of the system operators may be written as $\sigma_x (t) = \sigma_x$, and $\sigma_y(t) = \cos(\Omega_r t)\sigma_y + \sin(\Omega_r t)\sigma_z$. 
Using these expressions we find that the phonon contribution takes the compact form
\begin{equation}\begin{split}
\mathcal{K}_{{\rm {\rm PH}}}[\tilde\rho_S(t)] &= -\int\limits_0^\infty d\tau ~\text{tr}_B\left[H^{{\rm {\rm PH}}}_I(t),\left[H^{{\rm {\rm PH}}}_I(t-\tau), \tilde\rho_S(t)\otimes\rho_B^{{\rm PH}}\right]\right],\\
&= -\sum\limits_{j\in\{x,y,z\}}\left(\left[\sigma_x,\sigma_j\tilde\rho_S(t)\right]\Gamma_j + \left[\sigma_y,\sigma_j\tilde\rho_S(t)\right]\chi_j + \text{h.c.}\right).
\end{split}\end{equation}   
The transition rates induced by the phonon environment may be written as,
\begin{equation}\begin{split}
&\Gamma_x = \int\limits_0^\infty d\tau\Lambda_{xx}(\tau), \hspace{0.5cm}\chi_y = \int\limits_0^\infty d\tau\cos(\Omega_r\tau)\Lambda_{yy}(\tau),\hspace{0.5cm}\text{and}\hspace{0.5cm}\chi_z = \int\limits_0^\infty d\tau\sin(\Omega_r\tau)\Lambda_{yy}(\tau),
\end{split}
\end{equation}
with $\chi_x=\Gamma_y = \Gamma_z = 0$, where $\Lambda_{jj}(\tau) = \text{tr}_B( B_{j}(\tau)B_{j}\rho_B^{{\rm PH}})$ denotes the correlation function of the phonon environment in the polaron frame. 
By evaluating the trace over the phonon degrees of freedom, the correlation functions take the form~\cite{1367-2630-12-11-113042} $\Lambda_{xx}(\tau) =  B^2(\exp(\varphi(\tau)) + \exp(-\varphi(\tau)) - 2)/2$ and  $\Lambda_{yy}(\tau) =  B^2(\exp(\varphi(\tau)) - \exp(-\varphi(\tau)))/2$, where we have defined $\varphi(\tau) = \int_0^\infty\frac{J_{{\rm PH}}(\nu)}{\nu^2}(\coth\left(\frac{\beta\nu}{2}\right)\cos\nu\tau - i\sin\nu\tau)d\nu$. The coupling between the system and the environment is contained within the spectral density, which we take to have the standard form~\cite{nazir2008photons,PhysRevLett.104.017402s,PhysRevLett.105.177402s}, $J_{{\rm PH}}(\nu) = \alpha\nu\exp(-\nu^2/\nu_c^2)$. 
A detailed account of the validity of the polaron theory may be found in Refs.~\cite{1367-2630-12-11-113042s,nazir2015modellings}.

\subsection{The photon contribution}
We now focus on the interaction between the electromagnetic field and the QD. 
The interaction picture Hamiltonian for the field may be written as $H_I^{{\rm EM}} = \sigma^\dagger(t) e^{i\omega_l t} B_+(t) A(t)  + \text{h.c.}$, where $A(t) = \sum_m f_m {a_m} e^{-i\omega_m t}$ and $B_+(t)$ is as given in the previous section. 
If we consider the interaction picture transformation for the system operators we have
\begin{equation}
\sigma(t)e^{-i\omega_lt} = e^{i\frac{\Omega_r}{2}\sigma_x t }\sigma e^{-\frac{\Omega_r}{2}\sigma_x t} e^{-i\omega_l t}\approx \sigma e^{-i\omega_X^\prime t},
\end{equation}
where we have used the fact that $\omega_l\gg\Omega_r$ for typical QD systems, and that $\omega_l = \omega_X^\prime$ for resonant driving, to simplify the interaction picture transformation~\cite{carmichael1998statisticals,McCutcheon2013s}. 
By substituting this expression into the photon contribution of the master equation, and assuming all modes of the field are in their vacuum state, $\rho_B^{{\rm EM}} = \bigotimes_m\ket{0_m}\bra{0_m}$, we have
\begin{equation}
\mathcal{K}_{{\rm EM}}[\tilde\rho_s(t)] = -\int\limits_0^\infty d\tau ~\text{tr}_B\left[H^{{\rm EM}}_I(t),\left[H^{{\rm EM}}_I(t-\tau), \tilde\rho_S(t)\otimes\rho_B^{{\rm EM}}\right]\right] = \frac{\Gamma(\omega_X^\prime)}{2}\left(2\sigma\rho\sigma^\dagger-\left\{\sigma^\dagger\sigma,\tilde\rho_S(t)\right\}\right)
\end{equation}
where the spontaneous emission rate is given by~\cite{McCutcheon2013s,roy2015spontaneouss,PhysRevB.92.205406s}
\begin{equation}
\Gamma(\omega_X^\prime) = \text{Re}\left[\int\limits_0^\infty e^{i\omega^\prime_X\tau}G(\tau) \Lambda(\tau)d\tau\right].
\end{equation}
Here, $G(\tau) = \tr_B(B_\pm(\tau)B_\mp) = B^2e^{\varphi(t)}$ is the phonon correlation function and $\Lambda(\tau) = \int_0^\infty d\omega J_{{\rm EM}}(\omega)e^{i\omega\tau}$, with $J_{{\rm EM}}(\omega) = \sum_m\vert f_m\vert^2\delta(\omega - \omega_m)$ being the spectral density of the electromagnetic environment. 
As discussed in the manuscript, the local density of states of the electromagnetic field does not vary appreciably over energy scales relevant to QD systems in bulk, which allows us to make the assumption that the spectral density is flat~\cite{carmichael1998statisticals,McCutcheon2013s}, $J_{{\rm EM}}(\omega)\approx2\gamma/\pi$. 
The electromagnetic correlation function may then be evaluated as $\Lambda(\tau)\approx  \gamma\delta(\omega)+i\mathcal{P}[1/\tau]$, where $\mathcal{P}$ denotes the principal value integral.
Combining these expressions and resolving the remaining integral, we find that the spontaneous emission rate takes on the form $\Gamma(\omega^\prime_X) \approx \gamma$, where we have used the fact that $G(0) = 1$, such that
\begin{equation}
\mathcal{K}_{{\rm EM}}[\tilde\rho_S(t)] = \frac{\gamma}{2}\left(2\sigma\rho\sigma^\dagger-\left\{\sigma^\dagger\sigma,\tilde\rho_S(t)\right\}\right).
\end{equation}

\section{Phonon Effects in optical emission properties}
As discussed in the main manuscript, for a flat spectral density of the electromagnetic environment, the field operator describing the emission properties of the QD in the polaron frame may be written in the Heisenberg picture as $\hat{E}^{+}(t) \propto B_-(t)\tilde\sigma(t)$, where $\tilde{\sigma}(t) = \sigma(t) e^{-i\omega_l t}$. 
In the subsequent sections we shall discuss the consequences that the presence of the phonon displacement operators have on the first- and second-order optical emission correlation functions.

\subsection{First-order correlation function and the phonon sideband} 
Using the Wiener-Khinchin theorem~\cite{steck2007quantums,carmichael1998statisticals}, one can show that the steady state spectrum of a field may be written as $S(\omega) =\text{Re}[\lim_{t\rightarrow\infty}\int_0^\infty\langle\hat{E}^{-}(t)\hat{E}^{+}(t+\tau)\rangle e^{i\omega\tau}d\tau]$. 
Using the expression for the field operators defined above, the QD emission may thus be written as,
\begin{equation}
S(\omega)\propto \text{Re}\left[\lim_{t\rightarrow\infty}\int_0^\infty\langle\sigma^\dagger(t)B_+(t)\sigma(t+\tau)B_-(t+\tau)\rangle e^{i(\omega-\omega_l)\tau}d\tau\right].
\end{equation}
In its current form, the above correlation function is intractable as there are displacement operators present that act on the multi-mode phonon environment.
We can, however, factorise this correlation function by recognising that the optical and phonon contributions are associated with very different timescales.
Thus $g^{(1)}(\tau) = \lim_{t\rightarrow\infty}\langle\sigma^\dagger(t)B_+(t)\sigma(t+\tau)B_-(t+\tau)\rangle\approx G(-\tau) g_0(\tau)$, with $G(\tau)$ as defined above and $g_0(\tau)=\lim_{t\rightarrow\infty}\langle\sigma^\dagger(t)\sigma(t+\tau)\rangle$.
This factorisation also allows us to separate the emission spectrum into a contribution from the purely optical transition of the QD and a contribution corresponding to emission via the phonon sideband. 
We do this by writing the QD emission spectrum as $S(\omega) = S_{{\rm EM}}(\omega) + S_{{\rm PH}}(\omega)$, where
\begin{equation*}
S_{{\rm EM}}(\omega) = B^2 \text{Re}\left[\int_0^\infty g_0(\tau) e^{i(\omega-\omega_l)\tau}d\tau\right]
\hspace{0.25cm}\text{and}\hspace{0.25cm}
S_{{\rm PH}}(\omega) =  \text{Re}\left[\int_0^\infty(G(-\tau) - B^2)g_0(\tau) e^{i(\omega-\omega_l)\tau}d\tau\right].
\end{equation*}
From these expressions it is easy to show that the total light emitted through the phonon sideband is given by $\int_{-\infty}^\infty S_{{\rm PH}}(\omega) d\omega =\pi(1-B^2)g_0(0)$, and through the direct optical transition by $\int_{-\infty}^\infty S_{{\rm EM}}(\omega) d\omega =\pi B^2 g_0(0)$.
We can also simplify the expression for $S_{{\rm PH}}(\omega)$ more generally by recognising that $(G(-\tau) - B^2)$ tends rapidly to zero in comparison to optical timescales, which allows us to replace $g_0(\tau)\approx g_0(0)$, such that $S_{{\rm PH}}(\omega) =  \text{Re}\left[g_0(0) \int_0^\infty(G(-\tau) - B^2) e^{i(\omega-\omega_l)\tau}d\tau\right]$.
%

\subsection{Second-order correlation function and two-photon interference}
We now wish to consider the effect that phonon processes have on the results of a two-photon interference experiment.
The scenario we wish to consider is analogous to the set-up used by Proux~\emph{et al.} in Ref.~\cite{PhysRevLett.114.067401s}, where the emission from the QD enters a Mach-Zehnder interferometer.
One arm of the interferometer has a time delay which is much greater than the coherence time of the incident photon, preventing single photon interference.
Both arms are then recombined on a 50:50 beam splitter, which has output ports labelled $\hat{E}^{+}_3$ and $\hat{E}^{+}_4$.

In a Hong-Ou-Mandel (HOM) experiment the quantity of interest is the steady-state second-order correlation function $G_{HOM}^{(2)}(\tau)= \lim_{t\rightarrow\infty}\langle \hat{E}^{-}_3(t)\hat{E}^{-}_4(t+\tau)\hat{E}^{+}_4(t+\tau)\hat{E}^{+}_3(t)\rangle
$, which gives the conditional probability that after a photon is detected in mode $\hat{E}^{+}_3$ at time $t$, a second is detected detected in mode $\hat{E}^{+}_4$ at time $t+\tau$.
Two photon interference occurs whenever $G^{(2)}_{HOM}(0) =0$, such that photons arriving simultaneously at the beam splitter are bunched when they leave.

We can relate the output fields to the input using the standard beam splitter transformations, $\hat{E}^+_3=(\hat{E}^+_1 + \hat{E}^+_2)/\sqrt{2}$ and $\hat{E}^+_4=(\hat{E}^+_1 - \hat{E}^+_2)/\sqrt{2}$. 
Applying this transformation to the second-order correlation function, we have
\begin{equation}
G^{(2)}_{HOM}(\tau) = \frac{1}{4}\lim_{t\rightarrow\infty}\left\langle\left(\hat{E}^-_1(t) + \hat{E}^-_2(t)\right)
\left(\hat{E}^-_1(t+\tau) - \hat{E}^-_2(t+\tau)\right)
\left(\hat{E}^+_1(t+\tau) - \hat{E}^+_2(t+\tau)\right)
\left(\hat{E}^+_1(t) + \hat{E}^+_2(t)\right)
\right\rangle.
\end{equation}
For the case of a continuously driven QD we can simplify this expression significantly: the time delay in one arm of interferometer allows us to treat the incident field modes as uncorrelated and independent, such that cross-terms in the correlation function factorise into the corresponding input fields, e.g. $\langle \hat{E}^{-}_2(t)\hat{E}^{-}_1(t+\tau)\hat{E}^{+}_1(t+\tau)\hat{E}^{+}_2(t)\rangle\rightarrow \langle \hat{E}^{-}_2(t)\hat{E}^{+}_2(t)\rangle\langle \hat{E}^{-}_1(t+\tau)\hat{E}^{+}_1(t+\tau)\rangle$.
Since the fields are now independent but identical, we may drop the labels corresponding to specific inputs.
Thus, after some algebra we may write $G^{(2)}_{HOM}(\tau)$ as
\begin{equation}\label{eq:G2func}
\begin{split}
G^{(2)}_{HOM}(\tau)=&\frac{1}{2}\lim_{t\rightarrow\infty}\left\{\langle \hat{E}^{-}(t)\hat{E}^{-}(t+\tau)\hat{E}^{+}(t+\tau) \hat{E}^{+}(t)\rangle\right.\\
&\left.+2\operatorname{Re}\left[\langle \hat{E}^{+}(t)\rangle\left(\langle \hat{E}^{-}(t)\hat{E}^{-}(t+\tau)\hat{E}^{+}(t+\tau)\rangle-\langle \hat{E}^{-}(t)\hat{E}^{-}(t+\tau)E^{+}(t)\rangle\right)\right]\right.\\
&\left.-\vert\langle \hat{E}^{-}(t+\tau) \hat{E}^{+}(t)\rangle\vert^2-\vert\langle \hat{E}^{+}(t+\tau) \hat{E}^{+}(t)\rangle\vert^2+\langle \hat{E}^{-}(t) \hat{E}^{+}(t)\rangle^2\right\}.
\end{split}
\end{equation}
Now, substituting in the expression for the field operator of the QD emission, we have
\begin{equation}\label{eq:G2func}
\begin{split}
G_{HOM}^{(2)}(\tau)&\propto\frac{1}{2}\lim_{t\rightarrow\infty}\left\{\langle \sigma^\dagger(t) B_{+}(t)\sigma^\dagger(t+\tau)B_{+}(t+\tau)\sigma(t+\tau)B_{-}(t+\tau)\sigma(t)B_{-}(t)\rangle\right.\\
&+2\operatorname{Re}\left[\langle \sigma(t)B_{-}(t)\rangle\left(\langle \sigma^\dagger(t)B_{+}(t)\sigma^\dagger(t+\tau)B_{+}(t+\tau)\sigma(t+\tau)B_{-}(t+\tau)\rangle\right.\right.\\
&\left.\left.-\langle \sigma^\dagger(t)B_{+}(t)\sigma(t+\tau)B_{-}(t+\tau)\sigma(t)B_{-}(t)\rangle\right)\right]\\
&\left.-\vert\langle \sigma^\dagger(t+\tau)B_{+}(t+\tau) \sigma(t)B_{-}(t)\rangle\vert^2-\vert\langle\sigma(t+\tau)B_{-}(t+\tau) \sigma(t)B_{-}(t)\rangle\vert^2+\langle \sigma^\dagger\sigma(t)\rangle^2\right\},
\\
&\approx\frac{1}{2}\lim_{t\rightarrow\infty}\left\{\langle\sigma^\dagger(t)\sigma^\dagger(t+\tau)\sigma(t+\tau)\sigma(t)\rangle\right.\\
&+2\operatorname{Re}\left[ B^2\langle\sigma\rangle_{ss}\left(\langle \sigma^\dagger(t)\sigma^\dagger(t+\tau)\sigma(t+\tau)\rangle-\mathcal{G}(\tau)\langle \sigma^\dagger(t)\sigma(t+\tau)\sigma(t)\rangle\right)\right]\\
&\left.-\vert G(\tau)\vert^2 \vert\langle\sigma^\dagger(t+\tau)\sigma(t)\rangle\vert^2 -\vert C(\tau)\vert^2 \vert \langle\sigma(t+\tau)\sigma(t)\rangle\vert^2+\langle\sigma^\dagger\sigma\rangle^2_{ss}
\right\},
\end{split}
\end{equation}
where we have factorised the phonon and QD operators in the correlation function as before. 
Here, we have defined $C(\tau) = \langle B_\pm(\tau) B_\pm\rangle = B^2 e^{-\varphi(\tau)}$ and $\mathcal{G}(\tau) = B^{-1}\langle B_+ B_+(\tau) B_-\rangle = e^{\varphi(\tau) - \varphi^\ast(\tau)}$. 
Finally, it is convention to normalise this correlation function, such that
\begin{equation}
g^{(2)}(\tau) =  \lim_{t\rightarrow\infty}\frac{\langle \hat{E}^{-}_3(t)\hat{E}^{-}_4(t+\tau)\hat{E}^{+}_4(t+\tau)\hat{E}^{+}_3(t)\rangle}{\langle \hat{E}^-_3(t)\hat{E}^+_3(t)\rangle\langle \hat{E}^-_4(t)\hat{E}^+_4(t)\rangle} \propto\frac{G^{(2)}_{HOM}(\tau)}{\langle \sigma^\dagger\sigma\rangle_{ss}^2-\vert\langle \sigma\rangle_{ss}\vert^4},
\end{equation}
where $\langle\hat{O}\rangle_{ss}$ denotes the steady state expectation value of operator $\hat{O}$.

\providecommand{\noopsort}[1]{}\providecommand{\singleletter}[1]{#1}%

\end{document}